\documentclass[review]{elsarticle}
\usepackage{hyperref}
\usepackage[all]{hypcap}
\journal{Physics Letters A}
\usepackage{dsfont}
\usepackage{amsmath,amssymb}

\begin{document}

\begin{frontmatter}

\title{Tunneling in Polymer Quantization and the Quantum Zeno Effect}


\author[mysecondaryaddress]{Durmu{\c s} Ali Demir}
\ead{demir@physics.iztech.edu.tr}

\author[mysecondaryaddress]{Ozan Sarg{\i}n\corref{mycorrespondingauthor}}
\cortext[mycorrespondingauthor]{Corresponding author}
\ead{ozansargin@iyte.edu.tr}

\address[mysecondaryaddress]{Department of Physics, {\.I}zmir Institute of Technology, TR35430, {\.I}zmir, Turkey}

\begin{abstract}
As an application of the polymer quantization scheme, in this work we investigate the one dimensional quantum mechanical tunneling phenomenon from the perspective of polymer representation of a non-relativistic point particle and derive the transmission and reflection coefficients. Since any tunneling phenomenon inevitably evokes a tunneling time we attempt an analytical calculation of tunneling times by defining an operator well suited in discrete spatial geometry. The results that we come up with hint at appearance of the Quantum Zeno Effect in polymer framework.
\end{abstract}


\end{frontmatter}

\section{Introduction}
Quantizing gravity is probably the most challenging problem confronting today's theoretical physicists. Loop Quantum Gravity (LQG) is one of the formalisms pursued by physicists to reach this goal, and it has been quite successful in incorporating the background independent character demanded by general relativity. However, deep conceptual and practical differences between the background independent description and low energy description make it difficult to show that the former turns into the low energy description smoothly.\footnote{See \cite{Ashtekar2001} for an analysis in this direction.} That's exactly where the Polymer quantization comes into play. Polymer representation is a quantization scheme which is a low energy limit of LQG. It is a program initiated to inspect and resolve some conceptual problems in LQG in toy model settings.
\par
Polymer quantization approach has been used to study the features arising in loop quantum gravity \cite{Ashtekar2003,Willis2004,Fredenhagen2006} and especially loop quantum cosmology (LQC) \cite{Seahra,Kreienbuehl2013} since polymer framework and LQC have the same configuration space \cite{Velhinho2007}. This quantization scheme has been applied to toy models such as a free particle in one dimension \cite{Ashtekar2003,Willis2004} and simple harmonic oscillator \cite{Ashtekar2003,Willis2004,Hossain2010,BarberoG.2013}. \footnote{\cite{BarberoG.2013} stands out of the previous works on harmonic oscillator in that it conveys that the spectrum of the oscillator consists of bands similar to periodic potentials.} Galilean symmetries have been investigated \cite{Chiou2007}, continuum limit of polymer quantum systems has been explored \cite{Corichi2007,Corichi2007a}, singular potentials such as $1/r$ \cite{Husain2007} and $1/r^2$ \cite{Kunstatter2009} have been studied and it has been shown that polymer quantization leads to a modified uncertainty principle \cite{Hossain2010a}. Furthermore, statistical thermodynamics of a solid and ideal gas have been studied in \cite{Chacon-Acosta2011} and Bose-Einstein condensation has been investigated in \cite{Castellanos}. Entropies in the polymer  and standard Schr\"{o}dinger Hilbert spaces are analysed and they are shown to converge in the limit of vanishing polymer scale \cite{Demarie2013}.
\par
Tunneling is a purely quantum phenomenon that is caused by the uncertainty principle. Inspired by the fact that uncertainty principle gets modified in the framework of polymer quantum mechanics \cite{Hossain2010a}; in this work we study the tunneling of a non-relativistic quantum particle through a rectangular barrier in polymer quantization. The reason why we choose a rectangular barrier is that, one can decompose any type of potential into infinitesimal rectangular potential barriers.
\par
Tunneling time is the phenomenon that inevitably comes to one's mind along with tunneling. It has been bothering physicists for decades since the works of Condon in 1930 \cite{Condon1931} and MacColl in 1932 \cite{MacColl1932} for reasons that time is not represented by an operator in quantum mechanics \cite{Pauli1933} and classically tunneling time is imaginary \cite{Demir}.
Hence, we embark on  calculating tunneling times by defining a time operator which is odd when we consider the fact that time is just a parameter in quantum mechanics without an operator counterpart.
\par
This paper consists of three main parts. In the first, a review of the polymer particle representation is given. In the second, this quantization scheme is applied to the tunneling problem. Finally, in the last part, we consider tunneling times and arrive at rather interesting results regarding the validity of the Quantum Zeno effect in the framework of polymer quantization.
\section{Review of the Polymer Particle Description}
\subsection{Kinematics of the Polymer Representation}
In this section, we give a brief outline of the formulation and the notations of the polymer particle description; the details can be found in \cite{Ashtekar2003,Strocchi2005}.\par
According to the quantization procedure by Dirac, the first step in constructing a quantum theory out of a classical one is to define a quantization algebra, which replaces the Poisson bracket of observables in the classical theory.\par
In standard quantum theory, we employ Heisenberg algebra in which the position and  momentum operators satisfy CCRs
\begin{eqnarray}
[\hat{x},\hat{x}]=0 , \qquad [\hat{p},\hat{p}]=0 \qquad [\hat{x},\hat{p}]=i\hbar\,.
\end{eqnarray}
\par
In Polymer Quantization, however, one adopts the  Weyl algebra which is defined by the following Weyl relations.
\begin{itemize}
  \item $\hat{U}(\lambda)\hat{V}(\mu) = e^{-i\lambda\mu} \hat{V}(\mu)\hat{U}(\lambda)$
  \item $\hat{U}(\lambda_{1})\hat{U}(\lambda_{2})= \hat{U}(\lambda_{1} +\lambda_{2})$
  \item $\hat{V}(\lambda_{1})\hat{V}(\lambda_{2})= \hat{V}(\lambda_{1} +\lambda_{2})$.
\end{itemize}
If the one parameter unitary operators $U(\lambda)$ and $V(\mu)$ are weakly continuous in their parameters, Heisenberg algebra and Weyl algebra can be related to one another. Namely, we can write the unitary operators $\hat{U}(\lambda)$ and $\hat{V}(\mu)$ in terms of exponentiated position and momentum operators,
\begin{equation}
\hat{U}(\lambda)= e^{i\lambda\hat{x}}, \qquad \hat{V}(\mu)= e^{\frac{i\mu\hat{p}}{\hbar}}.
\end{equation}
In Schr\"{o}dinger quantum mechanics, the Hilbert space is Lebesgue measurable, this ensures that the weak continuity condition is satisfied. In polymer quantization, however, the kinematical Hilbert space is the Cauchy completion of cylindrical functions defined on a discrete topology. The discrete topology appears as a discrete inner product in the position basis, namely
\begin{equation}
\langle x_{i}|x_{j}\rangle=\delta_{i,j}.
\end{equation}
Since space is endowed with a discrete topology in polymer quantization, weak continuity condition is not satisfied for $\hat{V}(\mu)$, hence there is not a one to one correspondence between $\hat{V}(\mu)$ and $\hat{p}$. This is not surprising because, we do not expect momentum operator to exist in such a geometry since it is defined through differentiation.\par
So, in polymer scheme we have the discrete position operator and the one-parameter unitary operator $\hat{V}(\mu)$ at hand. The actions of these on the position basis are represented by
\begin{eqnarray}
\hat{x}|x_{j}\rangle = x_{j}|x_{j}\rangle 
\end{eqnarray}
\begin{eqnarray}\label{eight}
\hat{V}(\mu)|x_{j}\rangle = |x_{j}-\mu\rangle.
\end{eqnarray}
\subsection{Dynamics of Polymer Representation}
The analog of the Schr\"{o}dinger momentum operator is defined in this construction as $\hat{p}=\hbar\hat{K}_{\mu_{0}}$, where $\hat{K}_{\mu_{0}}=\frac{\hat{V}(-\mu_{0}) - \hat{V}(\mu_{0})}{-2i\mu_{0}}$ .
The generic classical Hamiltonian is of the form $H=\frac{p^2}{2m}+W(x)$ .
\par
Since $\hat{x}$ is well-defined, the main problem is that of defining the operator analog of $\hat{p}^2$ and thereby regularizing the Hamiltonian. For this purpose, we use the definition $\hat{p}=\hbar\hat{K}_{\mu_{0}}$ and obtain the Hamiltonian in terms of the shifting operator $\hat{V}(\mu)$ as
\begin{eqnarray}
\hat{H}_{\mu_{0}}=\frac{\hbar^2}{2m\mu_{0}^2}\left[2-\hat{V}(\mu_{0})-\hat{V}(-\mu_{0})\right]+\hat{W}(\hat{x}).
\end{eqnarray}
\par
The energy eigenvalue problem $\hat{H}_{\mu_{0}}\psi=E\psi$ takes the form of a second order difference equation in the position representation:
\begin{eqnarray}
\psi(x+\mu_{0})+\psi(x-\mu_{0})=\left[2-\frac{2m\mu_{0}^2}{\hbar^2}(E-W(x))\right]\psi(x).
\end{eqnarray}
\section{Tunneling in Polymer Representation}
In this section, we will investigate the tunneling problem using the polymer representation. The shape of the potential barrier is given below in figure \ref{potential}. As it is depicted in the figure, we study the problem by dividing the potential into three regions. In each region we solve the relevant eigenvalue equation and find the wave function in that region. In the end, we apply the boundary conditions and calculate the transmission and reflection coefficients. A remark is in order about the potential profile here: the barrier width is assumed to be $L=N\mu_{0}$, where $\mu_{0}$ is the fundamental length scale in polymer representation and $N$ is an integer.
\begin{figure}[h]\label{potential}
\centering
\includegraphics[scale=0.6]{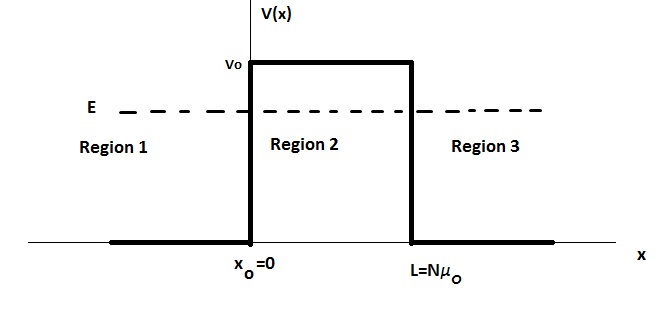}
\caption{Shape of the potential barrier}
\end{figure}
\subsection{\textbf{Region 1:}}
In this region, the Hamiltonian takes the form of a free particle, namely :
\begin{eqnarray}
\label{hamiltonian1}
\hat{H}_{1}=\frac{\hbar^2}{2m\mu_{0}^2}\left[2-\hat{V}(\mu_{0})-\hat{V}(-\mu_{0})\right].
\end{eqnarray}
\par
Using the fact that $x_{j}=x_{0}+j\mu_{0}$ and defining the state vectors in the polymer framework as $|\psi\rangle=\sum_{j\in\mathds{Z}}\psi(x_{j})|x_{j}\rangle$
the eigenvalue equation $\hat{H}_{\mu_{0}}|\psi\rangle=E|\psi\rangle$ takes the form:
\begin{eqnarray}
\label{eq25}
\frac{\hbar^2}{2m\mu_{0}^2} \Big[2\psi(x_{j})-\psi(x_{j}-\mu_{0})-\psi(x_{j}+\mu_{0}) \Big]=E\psi(x_{j}).
\end{eqnarray}
After making the redefinition $x_{j}\equiv j+1$ and doing the necessary manipulations, equation (\ref{eq25}) becomes:
\begin{eqnarray}
\psi(j+2)-\left(2-\frac{2mE\mu_{0}^2}{\hbar^2}\right)\psi(j+1)+\psi(j)=0\,.
\end{eqnarray}
The solution of this second order difference equation is proposed to be $\psi(j)=a_{+}r_{+}^j+a_{-}r_{-}^j$
where $a_{\pm}$ are constant coefficients and the roots of the characteristic equation $r^2-\left(2-\frac{2mE\mu_{0}^2}{\hbar^2}\right)r+1=0$, $r_{\pm}$ , are given by:
\begin{eqnarray}
\label{characteristic}
r_{\pm}=\left(1-\frac{mE\mu_{0}^2}{\hbar^2}\right)\pm\frac{1}{2}\sqrt{\frac{8mE\mu_{0}^2}{\hbar^2}\left(\frac{mE\mu_{0}^2}{2\hbar^2}-1\right)}\quad.
\end{eqnarray}
Equation (\ref{characteristic}) can be written in the following simpler form; $r_{\pm}=\varepsilon\pm\sqrt{\varepsilon^2-1}$
where $\varepsilon\equiv\left(1-\frac{mE\mu_{0}^2}{\hbar^2}\right)$.
In order to obtain physical solutions in \emph{Region 1} the roots of the characteristic equation should be complex numbers; that is $\varepsilon^2<1$. This leads to the idea that the minimum length scale $\mu_{0}$ imposes a cut-off on the energy; namely we should have $E<\frac{2\hbar^2}{m\mu_{0}^2}$ .
Using this fact, the wave function becomes 
\begin{eqnarray}
\psi(j)=a_{+}\left(\varepsilon+i\sqrt{1-\varepsilon^2}\right)^j+a_{-}\left(\varepsilon-i\sqrt{1-\varepsilon^2}\right)^j.
\end{eqnarray}
We can write this equation in polar coordinates by defining $\varepsilon\equiv\cos\theta$ and $\sqrt{1-\varepsilon^2}=\sin\theta$ .
The result becomes
\begin{eqnarray}
\psi_{1}(j)=a_{1}(\cos\theta+i\sin\theta)^{j}+a_{2}(\cos\theta-i\sin\theta)^{j}
\end{eqnarray}
which is equal to
\begin{eqnarray}
\label{wavefunction1}
\psi_{1}(j)=a_{1}e^{ij\theta}+a_{2}e^{-ij\theta}
\end{eqnarray}
where $\theta=\arccos(\varepsilon)$. Plugging in the values of $\theta$ and $\varepsilon$, equation (\ref{wavefunction1}) becomes
\begin{eqnarray}
\label{wavefunctionbeforemom}
\psi_{1}(j)=a_{1}e^{ij\arccos(1-\frac{m\mu_{0}^2}{\hbar^2}E)}+a_{2}e^{-ij\arccos(1-\frac{m\mu_{0}^2}{\hbar^2}E)}.
\end{eqnarray}
\par
\subsection{\textbf{Region 2:}}
In this region, the Hamiltonian and the eigenvalue equation $\hat{H}_{\mu_{0}}|\psi\rangle=E|\psi\rangle$ take respectively the forms:
\begin{eqnarray}
\label{hamiltonian2}
\hat{H}_{2}=\frac{\hbar^2}{2m\mu_{0}^2}\left[2-\hat{V}(\mu_{0})-\hat{V}(-\mu_{0})\right]+V_{0}
\end{eqnarray}
and
\begin{eqnarray}
\psi(j+2)-\left(2-\frac{2m(E-V_{0})\mu_{0}^2}{\hbar^2}\right)\psi(j+1)+\psi(j)=0. \nonumber
\end{eqnarray}

The characteristic equation corresponding to this difference equation is $r^2-\left(2-\frac{2m(E-V_{0})\mu_{0}^2}{\hbar^2}\right)r+1=0$.
The roots of this characteristic equation are $r_{\pm}=\lambda\pm\sqrt{\lambda^2-1}$
where $\lambda\equiv\left(1-\frac{m(E-V_{0})\mu_{0}^2}{\hbar^2}\right)$.
For real and distinct roots $\lambda^2>1$.
 In that case, the proposed solution of the difference equation takes the form $\psi_{2}(j)=b_{1}(\lambda+\sqrt{\lambda^2-1})^j+b_{2}(\lambda-\sqrt{\lambda^2-1})^j$.
 Which, by making the definition $\lambda\equiv\cosh\phi$, becomes $\psi_{2}(j)=b_{1}e^{j\phi}+b_{2}e^{-j\phi}$
where $\phi= {\rm arccosh} \left(1-\frac{m(E-V_{0})\mu_{0}^2}{\hbar^2}\right)$.
 Hence, the wave function in this region is
\begin{eqnarray}
\psi_{2}(j)=b_{1}e^{j{\rm arccosh} \left(1-\frac{m(E-V_{0})\mu_{0}^2}{\hbar^2}\right)}+b_{2}e^{-j{\rm arccosh} \left(1-\frac{m(E-V_{0})\mu_{0}^2}{\hbar^2}\right)}.
\end{eqnarray}
\subsection{\textbf{Region 3:}}
Region 3 has the same Hamiltonian as Region 1, namely equation (\ref{hamiltonian1}). The characteristic equation has the roots (\ref{characteristic}) and they are written compactly as $r_{\pm}=\varepsilon\pm\sqrt{\varepsilon^2-1}$ .
For complex roots, i.e. a physical wave function, we should have $\epsilon^2<1$ .
 The wave function, as in the first region , is $\psi_{3}(j)=c_{1}e^{ij\arccos\varepsilon}+c_{2}e^{-ij\arccos\varepsilon}$
but since on the right side of the barrier we should have only a right propagating wave, the coefficient $c_{2}$  of the second term must be zero.
Hence, the wave function reduces to
\begin{eqnarray}
\label{wavefunction31}
\psi_{3}(j)=c_{1}e^{ij\arccos\left(1-\frac{m\mu_{0}^2}{\hbar^2}E\right)}.
\end{eqnarray}
\subsection{\textbf{Transmission and Reflection Coefficients}}\par
Conservation of probability current dictates that we equate the wave-functions and their derivatives at the boundaries. At the left-end of the barrier, $\psi_{1}(0)=\psi_{2}(0)$ gives us
\begin{eqnarray}
\label{coef1}
a_{1}+a_{2}=b_{1}+b_{2}.
\end{eqnarray}
At the right-end $\psi_{2}(N)=\psi_{3}(N)$ returns
\begin{eqnarray}
\label{coef2}
b_{1}e^{N\rm arccosh(\lambda)}+b_{2}e^{-N\rm arccosh(\lambda)}=c_{1}e^{iN\arccos(\varepsilon)}.
\end{eqnarray}
 Derivatives of the wave function are calculated using the definition of derivative as a limit. Equating the derivatives of the wave function at the left and the right-ends, result in the following equations respectively:
%
\begin{equation}\label{coef3}
\begin{split}
a_{1}\left(1-e^{-i\arccos(\varepsilon)}\right)+a_{2}\left(1-e^{i\arccos(\varepsilon)}\right)= \,&b_{1}\left(e^{\rm arccosh(\lambda)}-1\right)+\\&b_{2}\left(e^{-\rm arccosh(\lambda)}-1\right)
\end{split}
\end{equation}

\begin{equation}\label{coef4}
\begin{split}
&b_{1}\left(e^{\rm arccosh(\lambda)}-1\right)e^{(N-1)\rm arccosh(\lambda)}+b_{2}\left(e^{-\rm arccosh(\lambda)}-1\right)e^{-(N-1)\rm arccosh(\lambda)}= \\
&c_{1}\left(e^{i\arccos(\varepsilon)}-1\right)e^{iN\arccos(\varepsilon)}.
\end{split}
\end{equation}
\par
 Solving equations (\ref{coef1}), (\ref{coef2}), (\ref{coef3}) and (\ref{coef4}) simultaneously we find the analytical expressions for the coefficients $a_{1}$, $a_{2}$, $b_{1}$ and $b_{2}$ in terms of the undetermined coefficient $c_{1}$:
\begin{equation}\label{coef11}
\begin{split}
 a_{1}&=\frac{c_{1}e^{iN\arccos(\varepsilon)}}{\left(e^{2\rm arccosh(\lambda)}-1\right)\left(e^{i\arccos(\varepsilon)}-e^{-i\arccos(\varepsilon)}\right)}\Big(e^{(3-N)\rm  arccosh(\lambda)}- \\
 &4e^{(2-N)\rm arccosh(\lambda)}+2e^{i\arccos(\varepsilon)+(2-N)\rm arccosh(\lambda)}-4e^{i\arccos(\varepsilon)+(1-N)\rm arccosh(\lambda)}+\\
 &e^{2i\arccos(\varepsilon)+(1-N)\rm arccosh(\lambda)}+4e^{(1-N)\rm arccosh(\lambda)}-2e^{i\arccos(\varepsilon)+N\rm arccosh(\lambda)}+\\
 &4e^{N\rm arccosh(\lambda)}-e^{(N-1)\rm arccosh(\lambda)}+4e^{i\arccos(\varepsilon)+(N+1)\rm arccosh(\lambda)}-\\
 &4e^{(N+1)\rm arccosh(\lambda)}-e^{2i\arccos(\varepsilon)+(N+1)\rm arccosh(\lambda)}\Big)
\end{split}
\end{equation}

\begin{equation}\label{coef22}
\begin{split}
 a_{2}&=\frac{c_{1}e^{iN\arccos(\varepsilon)}}{\left(e^{2\rm arccosh(\lambda)}-1\right)\left(e^{i\arccos(\varepsilon)}-e^{-i\arccos(\varepsilon)}\right)}\Big(
 -e^{i\arccos(\varepsilon)+(2-N)\rm arccosh(\lambda)}+ \\
 &2e^{i\arccos(\varepsilon)+(1-N)\rm arccosh(\lambda)}
 -2e^{i\arccos(\varepsilon)+(N+1)\rm arccosh(\lambda)}-5e^{(1-N)\rm arccosh(\lambda)}+\\
 &e^{i\arccos(\varepsilon)+N\rm arccosh(\lambda)}
 -e^{-i\arccos(\varepsilon)+(2-N)\rm arccosh(\lambda)}+e^{-i\arccos(\varepsilon)+N\rm arccosh(\lambda)}+\\
 &2e^{-i\arccos(\varepsilon)+(1-N)\rm arccosh(\lambda)}
 -2e^{-i\arccos(\varepsilon)+(N+1)\rm arccosh(\lambda)}
 +5e^{(N+1)\rm arccosh(\lambda)}\\
 &-e^{(3-N)\rm arccosh(\lambda)}
 +4e^{(2-N)\rm arccosh(\lambda)}
 -4e^{N\rm arccosh(\lambda)}
 +e^{(N-1)\rm arccosh(\lambda)}\Big)
\end{split}
\end{equation}

\begin{equation}\label{coef33}
\begin{split}
 b_{1}=\frac{\left(c_{1}e^{iN\arccos(\varepsilon)}\right)}{\left(e^{2\rm arccosh(\lambda)}-1\right)}\Big(&e^{(2-N)\rm arccosh(\lambda)}+e^{i\arccos(\varepsilon)+(1-N)\rm arccosh(\lambda)}-\\
 &2e^{(1-N)\rm arccosh(\lambda)}\Big)
\end{split}
\end{equation}

\begin{equation}\label{coef44}
\begin{split}
 b_{2}=\frac{\left(c_{1}e^{iN\arccos(\varepsilon)}\right)}{\left(e^{2\rm arccosh(\lambda)}-1\right)}\Big(&2e^{(N+1)\rm arccosh(\lambda)}-e^{i\arccos(\varepsilon)+(N+1)\rm arccosh(\lambda)}-\\
 &e^{N\rm arccosh(\lambda)}\Big)
\end{split}
\end{equation}
\par
The transmission and reflection coefficients are defined respectively as $T=\frac{|c_{1}|^2}{|a_{1}|^2}$ and $R=\frac{|a_{2}|^2}{|a_{1}|^2}$ .
When we plug  the values of $ a_{1}$  and $a_{2}$ in the transmission and reflection coefficients and sum them it is easily obtained that $T+R=1$, which is the requirement of probability conservation. \setlength{\footnotesep}{0.2mm}
\section{Tunneling time and Quantum Zeno Effect}
Time and position are not treated on an equal footing in quantum theory. Position is represented by an operator whereas time is left as just a parameter. Here, in our work, we end this dichotomy between time and position by elevating time to the status of being represented by an operator. First of all, we define our time-operator and then use it to calculate the time it takes for a quantum particle to tunnel through the potential barrier given in figure \ref{potential}.
\par
We define our differential time-operator as :
\begin{eqnarray}
\label{diftim}
d\hat{T}=\Big|\frac{m d\hat{x}}{\hat{p}}\Big|.
\end{eqnarray}
The reason why we define our differential time operator through an absolute value sign is that we have to make sure that the differential length, $d\hat{x}$, and momentum, $\hat{p}$, of the particle point in the same direction so that when we integrate out the differential time operator we get the tunneling time that corresponds to transmission.
When we use the regularized momentum operator, $\hat{p}=\frac{\hbar}{i2\mu_{0}}\left(\hat{V}(\mu_{0})-\hat{V}(-\mu_{0})\right)$, in (\ref{diftim}) we get:
\begin{eqnarray}
\label{diftime}
d\hat{T}=\Biggr|\left(\frac{i2m\mu_{0}}{\hbar}\right)\frac{d\hat{x}}{\hat{V}(\mu_{0})-\hat{V}(-\mu_{0})}\Biggr|.
\end{eqnarray}
\par
Tunneling-time is the expectation value of the integral of differential time-operator from the left-end of  the barrier, $x=0$, to the right-end, $x=L$. Hence, we can write tunneling time as:
\begin{eqnarray}
\label{time}
T=\langle\psi|\Biggr|\left(\int_{0}^{L} \left(\frac{i2m\mu_{0}}{\hbar}\right)\frac{d\hat{x}}{\hat{V}(\mu_{0})-\hat{V}(-\mu_{0})} \right)\Biggr||\psi\rangle\,
\end{eqnarray}
In our analysis we are going to work in the position basis since we have a better knowledge of the position of the particle during the tunneling process. But in order to do that, we have to regularize the momentum operator somehow and bring the expression in the denominator of the time operator, i.e. $\hat{V}(\mu_{0})-\hat{V}(-\mu_{0})$ , up into the nominator because position eigenkets are not eigenkets of the operator $\hat{V}(\mu_{0})$. For this purpose, we will use the regularization
\begin{eqnarray}
\label{shiftmom}
\frac{1}{\hat{V}(\mu_{0})-\hat{V}(-\mu_{0})}=\int_{0}^{\infty} e^{-a\left(\hat{V}(\mu_{0})-\hat{V}(-\mu_{0})\right)} da
\end{eqnarray}
and then employ the series expansion formula
for the exponential. After doing this, we plug (\ref{shiftmom}) into (\ref{time}) and insert the identity operators between the wave functions and the time operator. The result becomes:
\begin{equation}
\begin{split}
T=&\Biggr|\left(\frac{i2mN\mu_{0}^2}{\hbar}\right)\sum_{k}\int_{0}^{\infty}da\Big[\Big(|b_{1}|^2 e^{2k\rm arccosh(\lambda)}+b_{1}^{*}b_{2}+|b_{2}|^2 e^{-2k\rm arccosh(\lambda)}+\\
&b_{2}^{*}b_{1}\Big)-\frac{a}{1!}\left(e^{-\rm arccosh(\lambda)}-e^{\rm arccosh(\lambda)}\right)\Big(|b_{1}|^2 e^{2k\rm arccosh(\lambda)}+b_{1}^{*}b_{2}-b_{2}^{*}b_{1}-\\
&|b_{2}|^2 e^{-2k\rm arccosh(\lambda)}\Big)
+\frac{a^2}{2!}\left(e^{\rm arccosh(\lambda)}-e^{-\rm arccosh(\lambda)}\right)^2\Big(|b_{1}|^2 e^{2k\rm arccosh(\lambda)}+\\
&b_{1}^{*}b_{2}+b_{2}^{*}b_{1}+|b_{2}|^2 e^{-2k\rm arccosh(\lambda)}\Big)
-\frac{a^3}{3!}\left(e^{-\rm arccosh(\lambda)}-e^{\rm arccosh(\lambda)}\right)^3\times\\
&\left(|b_{1}|^2 e^{2k\rm arccosh(\lambda)}+b_{1}^{*}b_{2}-b_{2}^{*}b_{1}-|b_{2}|^2 e^{-2k\rm arccosh(\lambda)}\right)+\ldots\Big]\Biggr|\;\,.
\end{split}
\end{equation}
After a little bit of algebra this equation can be recast in the form
\begin{equation}\label{time2}
\begin{split}
T=&\Biggr|\left(\frac{i2mN\mu_{0}^2}{\hbar}\right)\int_{0}^{\infty}da\Biggr\{\sum_{k}\Big(|b_{1}|^2 e^{2k\rm arccosh(\lambda)}+b_{1}^{*}b_{2}+b_{2}^{*}b_{1}+\\
&|b_{2}|^2 e^{-2k\rm arccosh(\lambda)}\Big)\sum_{n=0}^{\infty}\Big[\frac{(2a)^{2n}(\lambda^2-1)^{n}}{(2n)!}\Big]
+\sum_{k}\Big(|b_{1}|^2 e^{2k\rm arccosh(\lambda)}+\\
&b_{1}^{*}b_{2}-b_{2}^{*}b_{1}-|b_{2}|^2 e^{-2k\rm arccosh(\lambda)}\Big)\sum_{n=0}^{\infty}\Big[\frac{(2a)^{2n+1}(\lambda^2-1)^{n+\frac{1}{2}}}{(2n+1)!}\Big]\Biggr\}\Biggr|.
\end{split}
\end{equation}
The two summations over the dummy index \emph{n} in the first and second terms of this equation are equal to $\cosh(2a\sqrt{\lambda^2-1})$ and $\sinh(2a\sqrt{\lambda^2-1})$, respectively. After collecting the terms, (\ref{time2}) takes the following form:
\begin{equation}\label{time3}
\begin{split}
T=\Biggr|\left(\frac{i2mN\mu_{0}^2}{\hbar}\right)\int_{0}^{\infty}da&\Biggr\{\sum_{k}\left(b_{2}^{*}b_{1}+|b_{2}|^2 e^{-2k\rm arccosh(\lambda)}\right)e^{-2a\sqrt{\lambda^2-1}}\\
&+\sum_{k}\left(|b_{1}|^2 e^{2k\rm arccosh(\lambda)}+b_{1}^{*}b_{2}\right)e^{2a\sqrt{\lambda^2-1}}\Biggr\}\Biggr|.
\end{split}
\end{equation}
The reader with a keen eye may have already noticed that the second exponential integral in (\ref{time3}) diverges. We will omit the diverging second term of this equation on the physical grounds that ($a\rightarrow\infty$)  corresponds to the zero momentum states and zero momentum inside the barrier amounts to no tunneling and hence to infinite tunneling time. To make this point clearer, one should revisit (\ref{shiftmom}) and  then realize that it is practically $\frac{1}{\hat{p}}=\int_{0}^{a_{max}} e^{-a\hat{p}} \quad da$ and inserting $\hat{p}=0$ in this equation corresponds to $a_{max}=\infty$. Since a zero momentum particle does not tunnel  through the barrier we can safely omit divergent parts of the tunneling time expression corresponding to those states.
After these comments, the equation we end up with is
\begin{equation}\label{time4}
\begin{split}
T=\Biggr|\left(\frac{i2mN\mu_{0}^2}{\hbar}\right)\int_{0}^{\infty} e^{-2a\sqrt{\lambda^2-1}} \quad da\Biggr\{\sum_{k=0}^{N}\left(b_{2}^{*}b_{1}+|b_{2}|^2 e^{-2k\rm arccosh(\lambda)}\right)\Biggr\}\Biggr|.\\
\end{split}
\end{equation}
After taking the integral and doing the summation this equation becomes;
\begin{equation}\label{time5}
\begin{split}
T=\Biggr|\left(\frac{i2mN\mu_{0}^2}{\hbar}\right)\left(\frac{1}{2\sqrt{\lambda^2-1}}\right)\Biggr\{&\left(b_{2}^{*}b_{1}\right)\left(1+N\right)+\\ &|b_{2}|^2 \frac{\left(e^{2\rm arccosh(\lambda)}-e^{-2N\rm arccosh(\lambda)}\right)}{\left(e^{2\rm arccosh(\lambda)}-1\right)}\Biggr\}\Biggr|.
\end{split}
\end{equation}
The next steps are  inserting the expressions for $b_{1}$ and  $b_{2}$; expressing $N$ in terms of the barrier length and $\mu_{0}$ and then finally using the relevant expressions for $\lambda$ and $\varepsilon$, which will come along with the terms $b_{2}^{*}b_{1}$ and $|b_{2}|^2$ . The explicit forms of these two parameters are, as we have stated before, $\lambda=\left(1-\frac{m(E-V_{0})\mu_{0}^2}{\hbar^2}\right)$ and $\varepsilon=\left(1-\frac{mE\mu_{0}^2}{\hbar^2}\right)$.
The outcome of all these operations is a rather lengthy time expression which we will not write out here.
\par
The next concept that we want to consider is the relation of this tunneling time to the Quantum Zeno Effect (QZE). QZE is a phenomenon in which a particular process may be slowed down or stopped as a result of frequent measurements and for this reason it gathers a lot of interest among physicists. Here we will not dwell  too deeply on the details of QZE since there is a good deal of  literature out there that may be consulted such as \cite{Pascazio2014,Facchi2008,Venugopalan2007,Alter1997,Sanz2012,Facchi2010,Peres1980,Facchi2010a} but let us  briefly articulate the physical content of this rather strange phenomenon. QZE may be defined simply as the inhibition of a quantum system's time evolution by frequent measurements of the system's state. This means that by frequent measurements you restrain the system from making a transition from an initial state to a final state and  in a sense , in the limit of infinitely many measurements in a finite time interval, the system's evolution is confined in a small subspace of the Hilbert space. The principle idea that led  Misra and Sudarshan in \cite{Misra1977} to predict the viability of the Quantum Zeno Effect is that unstable quantum systems were expected to exhibit a short-time non-exponential decay law. It was realized that, contrary to the classical heuristic exponential decay law, quantum systems follow three distinct  decay phases. The short-time phase is a quadratic one, the intermediate phase is the exponential decay and the long-time phase follows a power law. Misra and Sudarshan proposed that if frequent measurements are made in the short-time phase of the decay and if these measurements are ideal in the sense that they are von Neumann measurements which are represented by one-dimensional projectors; after each measurement the state of the system is projected back to the initial state, as a result time evolution of the system is slowed down and eventually comes to a halt. In recent years, it has been predicted that one may observe an increase in the decay rates of unstable systems as a result of frequent measurements if the frequency of observations is properly adjusted and in literature this phenomenon is referred to  as the Quantum Anti-Zeno Effect (\textbf{AZE}) \cite{Balachandran2000,Kofman2000}. The validity of Quantum Zeno and anti-Zeno effects are now both established. Experimental evidence for the non-exponential decay in quantum tunneling was reported in \cite{Wilkinson1997}, AZE and QZE are experimentally confirmed in works like \cite{Itano1990,Kwiat1995,Kofman1996,Zheng2013,Wolters2013}. In \cite{Fischer2001} Quantum Zeno and anti-Zeno effects are simultaneously reported to be observed.
\par
In works on QZE and AZE, it is stated that decay rates depend on the frequency of the measurements made on the system \cite{Facchi2001}. If the frequency of the observations is such that you measure the system's state each and every time  in the short-time phase of the decay then the decay rate drops; on the contrary, if the measurement frequency is such that you observe the system right at the point where the decay rate changes character , i.e. the point where the short-time phase gives way to the exponential phase, the decay accelerates.
\par
\begin{figure}[h!]
\centering
\includegraphics[scale=0.6]{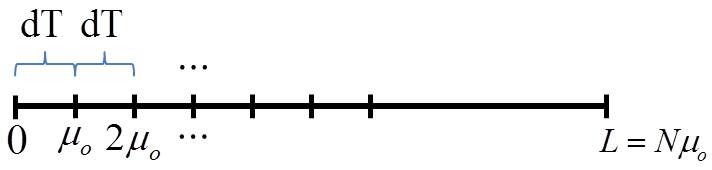}
\caption{Schematic illustration of connection between the polymeric length scale and frequency of position measurements.}\label{Fignumber36}
\end{figure}

In our work, the tunneling particle constitutes an unstable system and one may expect to observe Quantum Zeno and anti-Zeno effects through the change in the tunneling times as we change the characteristic length scale $\mu_{0}$. Altering the characteristic length scale amounts to altering the frequency of position measurements therefore changing the number of discrete steps the particle takes inside the barrier should effect the tunneling time in accordance with the Quantum Zeno and anti-Zeno effects. Let's make this point clearer.
Equation \ref{diftim} defines the differential tunneling time as $d\hat{T}=|\frac{md\hat{x}}{\hat{p}}|$.  We have used a regular lattice structure for space and this leads to the fact that at each step of the way through the barrier the particle traverses a constant length of $dx=\mu_{0}$. This in turn brings about a direct proportionality such that $dT\sim\mu_{0}$. It can be seen in the figure  \ref{Fignumber36} below that after each step of length $\mu_{0}$ we are effectively making a position measurement and also an amount of time $dT$ is added to the total tunneling time.
\par
\par
Inverting the proportionality $dT\sim\mu_{0}$, we see that the frequency of position measurements is inversely proportional to the fundamental length scale $\mu_{0}$ of polymer quantization, that is
\begin{equation}\label{eq1}
f\sim\frac{1}{\mu_{0}}.
\end{equation}
\par
Equation  (\ref{eq1}) leads to the fact that decreasing the fundamental length scale amounts to increasing the frequency of position measurements and this should result in the appearance of Quantum Zeno and Anti-Zeno effects in our work.

 We did a numerical analysis on our tunneling time expression, i.e. (\ref{time5}), to see if it accords with QZE and AZE claims made above. In that analysis, we expanded (\ref{time5}) in a Maclaurin Series in $\mu_{0}$ since it exquisitely depends on it. We have used $L=1nm$ as the barrier width, the tunneling particle is taken to be an electron, the height of the potential is taken to be $9.7eV$ and the energy of the electron $5.5eV$. The following figure , i.e. figure  \ref{Fignumber37}, is the result of this analysis. The solid curve in that figure depicts the general trend of our tunneling time expression with respect to changes in the fundamental length scale. Close inspection of this trend reveals that as we decrease the characteristic length scale $\mu_{0}$, i.e. increase the frequency of position measurements, tunneling time decreases up to a point and then as we continue to even smaller length scales the tunneling time displays a dramatic increase. We may interpret the part of the tunneling time curve that descends as we decrease $\mu_{0}$ as the anti-Zeno region and the other part where the tunneling time increases dramatically can be coined the Zeno region. The red  dashed vertical lines in the same figure correspond to the limits of the fundamental length scale for which the tunneling time is of the order of \textit{Femto seconds} which is the time scale of tunneling one encounters in some experiments in the literature like \emph{Steinberg, Kwiat, and Chiao (1993)} \cite{Steinberg1993}. We can read from this graph that the polymerization  scale is restricted to the interval between the red vertical lines, which extends from about \textit{9 pico meters} to about \textit{1.4 Angstrom}. Even though we are able to place limits on the polymerization scale by comparing the tunneling time estimates of our approach and experiments related to tunneling time, we have to emphasize that the potential profiles of the experimental set-ups one encounters in the literature related to tunneling times are in no way simple rectangular potentials like the one that we have employed in our work. Since our potential profile is dissimilar to the experimental ones, we should keep in mind that  limits we have placed on $\mu_{0}$ are not exact; on the contrary just crude ones.
\par
\begin{figure}[h!]
\label{Fignumber37}
\centering
\includegraphics[scale=0.8]{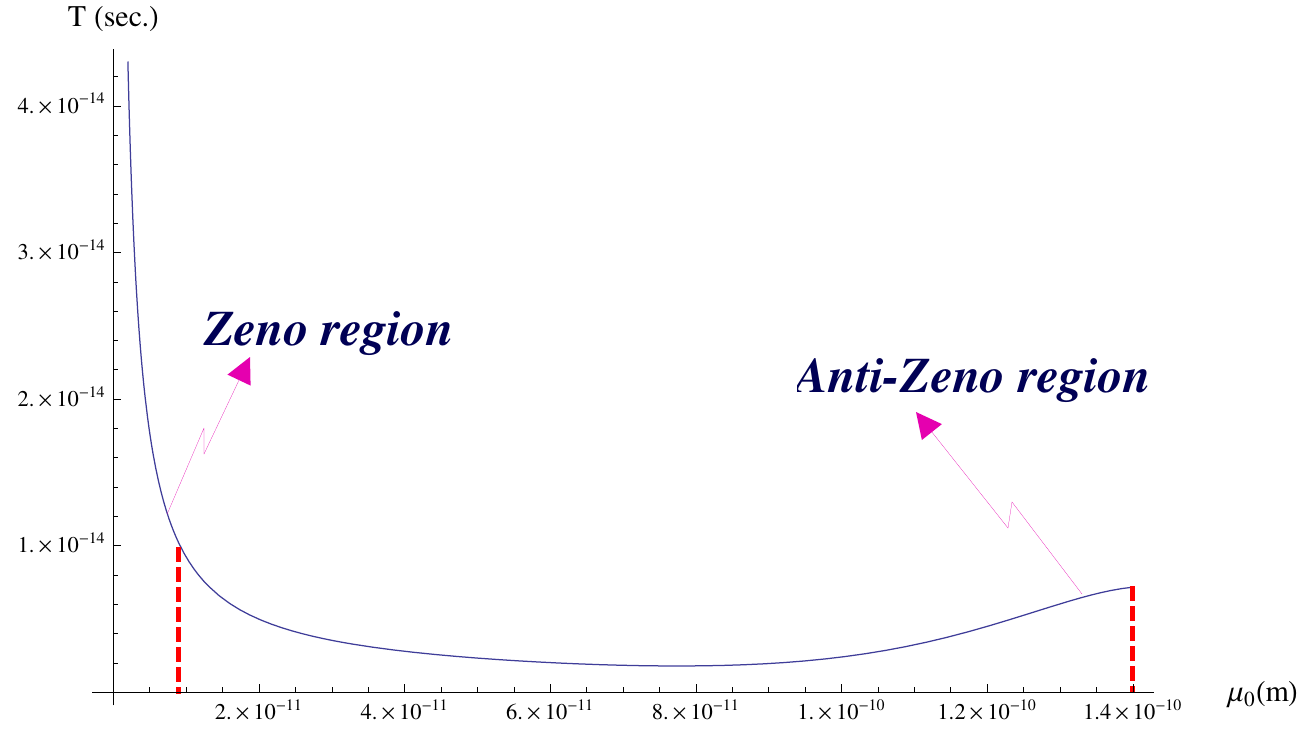}
\caption{Tunneling time against polymeric length scale. Red dashed lines mark the bounds of the polymerization scale imposed by the experiments.}
\end{figure}
\par
\section{Summary}
   We applied the polymer quantization formalism to the well known quantum tunneling phenomenon in order to see if it is possible to get sensible results similar to Schr\"{o}dinger formulation. For this purpose, we made use of a non-relativistic quantum particle in one dimension tunneling through a rectangular potential barrier. Since there is no counterpart to the Schr\"{o}dinger momentum operator in this scheme we had to introduce a new one in terms of shifting operator $\hat{V}(\mu)$. After regularizing the Hamiltonian using this operator, the eigenvalue equations pertaining to different regions of the potential turned into second order difference equations. Solutions of these gave us the wave functions. Conserving the probability current led to the four unknown coefficients out of the five. And then, the transmission and reflection coefficients were calculated and their sum, which is a consistency check on the formalism, was seen to be one as expected. Then, we defined a differential time operator to calculate the time it takes a non-relativistic particle to tunnel through the barrier. We calculated the tunneling time as the expectation value of the finite time operator, which is the integral of differential time operator between the boundaries of the barrier. Our calculations revealed that the tunneling time expression we got complies with the predictions of Quantum Zeno and anti-Zeno effects. The variation of the tunneling time with the fundamental length scale of polymer quantization reveals that as we decrease the length scale, thereby increase the discrete position steps that the particle takes and in a sense increase the number of position measurements made on the particle, the tunneling time first decreases up to a point and then as we continue to further decrease the length scale the tunneling time starts to increase dramatically. The part of the tunneling time curve to the right of the minimum of the curve can be identified with the behaviour inline with the Quantum anti-Zeno effect and the part to left can be coined the Quantum Zeno region.
\par

\section{Acknowledgements}
 We are grateful to Tu\u{g}rul G\"{u}ner for fruitful discussions.

\bibliography{makale}

\begin{thebibliography}{10}
\expandafter\ifx\csname url\endcsname\relax
  \def\url#1{\texttt{#1}}\fi
\expandafter\ifx\csname urlprefix\endcsname\relax\def\urlprefix{URL }\fi
\expandafter\ifx\csname href\endcsname\relax
  \def\href#1#2{#2} \def\path#1{#1}\fi

\bibitem{Ashtekar2001}
A.~Ashtekar, J.~Lewandowski, Relation between polymer and fock excitations,
  Classical and Quantum Gravity 18~(18) (2001) L117.

\bibitem{Ashtekar2003}
A.~Ashtekar, S.~Fairhurst, J.~L. Willis, Quantum gravity, shadow states and
  quantum mechanics, Classical and Quantum Gravity 20~(6) (2003) 1031,
  arXiv:gr-qc/0207106v3.

\bibitem{Willis2004}
J.~L. Willis, On the low-energy ramifications and a mathematical extension of
  loop quantum gravity, Ph.D. thesis, The Pennsylvania State University, The
  Graduate School, Eberly College of Science (2004).

\bibitem{Fredenhagen2006}
K.~Fredenhagen, F.~Reszewski, Polymer state approximation of schr\"{o}dinger
  wavefunctions, Classical and Quantum Gravity 23~(22) (2006) 6577,
  arXiv:gr-qc/0606090v2.

\bibitem{Seahra}
S.~S. Seahra, I.~A. Brown, G.~M. Hossain, V.~Husain, Primordial polymer
  perturbations,\; arXiv:1207.6714v2 [astro-ph.CO].

\bibitem{Kreienbuehl2013}
A.~Kreienbuehl, T.~Paw\l{}owski, Singularity resolution from polymer quantum
  matter, Phys. Rev. D 88 (2013) 043504, arXiv:1302.6566v2 [gr-qc].
\newblock \href {http://dx.doi.org/10.1103/PhysRevD.88.043504}
  {\path{doi:10.1103/PhysRevD.88.043504}}.

\bibitem{Velhinho2007}
J.~M. Velhinho, The quantum configuration space of loop quantum cosmology,
  Classical and Quantum Gravity 24~(14) (2007) 3745, arXiv:0704.2397v2 [gr-qc].

\bibitem{Hossain2010}
G.~M. Hossain, V.~Husain, S.~S. Seahra, Propagator in polymer quantum field
  theory, Phys. Rev. D 82 (2010) 124032.
\newblock \href {http://dx.doi.org/10.1103/PhysRevD.82.124032}
  {\path{doi:10.1103/PhysRevD.82.124032}}.

\bibitem{BarberoG.2013}
J.~F. {Barbero G.}, J.~Prieto, E.~J.~S. Villase{\~n}or, Band structure in the
  polymer quantization of the harmonic oscillator, Classical and Quantum
  Gravity 30~(16) (2013) 165011, arXiv:1305.5406v1 [gr-qc].

\bibitem{Chiou2007}
D.-W. Chiou, Galileo symmetries in polymer particle representation, Classical
  and Quantum Gravity 24~(10) (2007) 2603, arXiv:gr-qc/0612155v3.

\bibitem{Corichi2007}
A.~Corichi, T.~Vuka\v{s}inac, J.~A. Zapata, Polymer quantum mechanics and its
  continuum limit, Phys. Rev. D 76 (2007) 044016, arXiv:0704.0007v2 [gr-qc].
\newblock \href {http://dx.doi.org/10.1103/PhysRevD.76.044016}
  {\path{doi:10.1103/PhysRevD.76.044016}}.

\bibitem{Corichi2007a}
A.~Corichi, T.~Vuka\v{s}inac, J.~A. Zapata, Hamiltonian and physical hilbert
  space in polymer quantum mechanics, Classical and Quantum Gravity 24~(6)
  (2007) 1495, arXiv:gr-qc/0610072v2.

\bibitem{Husain2007}
V.~Husain, J.~Louko, O.~Winkler, Quantum gravity and the coulomb potential,
  Phys. Rev. D 76 (2007) 084002, arXiv:0707.0273v2.
\newblock \href {http://dx.doi.org/10.1103/PhysRevD.76.084002}
  {\path{doi:10.1103/PhysRevD.76.084002}}.

\bibitem{Kunstatter2009}
G.~Kunstatter, J.~Louko, J.~Ziprick, Polymer quantization, singularity
  resolution, and the $1/r^{2}$ potential, Phys. Rev. A 79 (2009) 032104,
  arXiv:0809.5098v2.
\newblock \href {http://dx.doi.org/10.1103/PhysRevA.79.032104}
  {\path{doi:10.1103/PhysRevA.79.032104}}.

\bibitem{Hossain2010a}
G.~M. Hossain, V.~Husain, S.~S. Seahra, Background-independent quantization and
  the uncertainty principle, Classical and Quantum Gravity 27~(16) (2010)
  165013, arXiv:1003.2207v1 [gr-qc].

\bibitem{Chacon-Acosta2011}
G.~Chac\'{o}n-Acosta, E.~Manrique, L.~Dagdug, H.~A. Morales-T\'{e}cotl,
  Statistical thermodynamics of polymer quantum systems, SIGMA 7 (2011) 110,
  arXiv:1109.0803v2 [gr-qc].

\bibitem{Castellanos}
E.~Castellanos, G.~Chac\'{o}n-Acosta, Polymer bose--einstein
  condensatesArXiv:1301.5362v1 [gr-qc].

\bibitem{Demarie2013}
T.~F. Demarie, D.~R. Terno, Entropy and entanglement in polymer quantization,
  Classical and Quantum Gravity 30~(13) (2013) 135006, arXiv:1209.3087v2
  [gr-qc].

\bibitem{Condon1931}
E.~U. Condon, P.~M. Morse, Quantum mechanics of collision processes i.
  scattering of particles in a definite force field, Rev. Mod. Phys. 3 (1931)
  43--88.
\newblock \href {http://dx.doi.org/10.1103/RevModPhys.3.43}
  {\path{doi:10.1103/RevModPhys.3.43}}.

\bibitem{MacColl1932}
L.~A. MacColl, Note on the transmission and reflection of wave packets by
  potential barriers, Phys. Rev. 40 (1932) 621--626.
\newblock \href {http://dx.doi.org/10.1103/PhysRev.40.621}
  {\path{doi:10.1103/PhysRev.40.621}}.

\bibitem{Pauli1933}
W.~Pauli, Handbuch der Physik, Vol.~24, Springer, Berlin, 1933.

\bibitem{Demir}
D.~A. Demir, Real-time tunneling\quad arXiv:quant-ph/9809036v2.

\bibitem{Strocchi2005}
F.~Strocchi, An Introduction to the Mathematical Structure of Quantum
  Mechanics, Vol.~27 of Advanced Series in Mathematical Physics, World
  Scientific Publishing Co. Pte. Ltd., 2005.

\bibitem{Pascazio2014}
S.~Pascazio, All you ever wanted to know about the quantum zeno effect in 70
  minutes \;, Open Systems \& Information Dynamics 21~(01n02) (2014) 1440007,
  arXiv:1311.6645v1 [quant-ph].
\newblock \href {http://dx.doi.org/10.1142/S1230161214400071}
  {\path{doi:10.1142/S1230161214400071}}.

\bibitem{Facchi2008}
P.~Facchi, S.~Pascazio, Quantum zeno dynamics: mathematical and physical
  aspects, J. Phys. A 41~(49) (2008) 493001, arXiv:0903.3297v1 [math-ph].

\bibitem{Venugopalan2007}
A.~Venugopalan, The quantum zeno effect — watched pots in the quantum world,
  Resonance 12~(4) (2007) 52--68, arXiv:1211.3498v1 [physics.hist-ph].
\newblock \href {http://dx.doi.org/10.1007/s12045-007-0038-x}
  {\path{doi:10.1007/s12045-007-0038-x}}.

\bibitem{Alter1997}
O.~Alter, Y.~Yamamoto, Quantum zeno effect and the impossibility of determining
  the quantum state of a single system, Phys. Rev. A 55 (1997) R2499--R2502.
\newblock \href {http://dx.doi.org/10.1103/PhysRevA.55.R2499}
  {\path{doi:10.1103/PhysRevA.55.R2499}}.

\bibitem{Sanz2012}
A.~Sanz, C.~Sanz-Sanz, T.~González-Lezana, O.~Roncero, S.~Miret-Artés,
  Quantum zeno effect: Quantum shuffling and markovianity, Ann. Phys. 327~(4)
  (2012) 1277 -- 1289, arXiv:1112.3829v2 [quant-ph].
\newblock \href {http://dx.doi.org/http://dx.doi.org/10.1016/j.aop.2011.12.012}
  {\path{doi:http://dx.doi.org/10.1016/j.aop.2011.12.012}}.

\bibitem{Facchi2010}
P.~Facchi, S.~Graffi, M.~Ligabò, The classical limit of the quantum zeno
  effect, J. Phys. A 43~(3) (2010) 032001, arXiv:0911.5675v1 [quant-ph].

\bibitem{Peres1980}
A.~Peres, Zeno paradox in quantum theory, Am. J. Phys. 48~(11) (1980) 931--932.
\newblock \href {http://dx.doi.org/http://dx.doi.org/10.1119/1.12204}
  {\path{doi:http://dx.doi.org/10.1119/1.12204}}.

\bibitem{Facchi2010a}
P.~Facchi, M.~Ligab\`{o}, Quantum zeno effect and dynamics, J. Math. Phys.
  51~(2) (2010) --, arXiv:0911.2201v1 [math-ph].
\newblock \href {http://dx.doi.org/http://dx.doi.org/10.1063/1.3290971}
  {\path{doi:http://dx.doi.org/10.1063/1.3290971}}.

\bibitem{Misra1977}
B.~Misra, E.~C.~G. Sudarshan, The zeno’s paradox in quantum theory, J. Math.
  Phys. 18~(4) (1977) 756--763.
\newblock \href {http://dx.doi.org/http://dx.doi.org/10.1063/1.523304}
  {\path{doi:http://dx.doi.org/10.1063/1.523304}}.

\bibitem{Balachandran2000}
A.~P. Balachandran, S.~M. Roy, Quantum anti-zeno paradox, Phys. Rev. Lett. 84
  (2000) 4019--4022, arXiv:quant-ph/9909056v1.
\newblock \href {http://dx.doi.org/10.1103/PhysRevLett.84.4019}
  {\path{doi:10.1103/PhysRevLett.84.4019}}.

\bibitem{Kofman2000}
A.~G. Kofman, G.~Kurizki, Acceleration of quantum decay processes by frequent
  observations, Nature 405~(6786) (2000) 546--550, arXiv:quant-ph/0102002v2.

\bibitem{Wilkinson1997}
S.~R. Wilkinson, C.~F. Bharucha, M.~C. Fischer, K.~W. Madison, P.~R. Morrow,
  Q.~Niu, B.~Sundaram, M.~G. Raizen, Experimental evidence for non-exponential
  decay in quantum tunnelling, Nature 387~(6633) (1997) 575--577.

\bibitem{Itano1990}
W.~M. Itano, D.~J. Heinzen, J.~J. Bollinger, D.~J. Wineland, Quantum zeno
  effect, Phys. Rev. A 41 (1990) 2295--2300.
\newblock \href {http://dx.doi.org/10.1103/PhysRevA.41.2295}
  {\path{doi:10.1103/PhysRevA.41.2295}}.

\bibitem{Kwiat1995}
P.~Kwiat, H.~Weinfurter, T.~Herzog, A.~Zeilinger, M.~A. Kasevich,
  Interaction-free measurement, Phys. Rev. Lett. 74 (1995) 4763--4766.
\newblock \href {http://dx.doi.org/10.1103/PhysRevLett.74.4763}
  {\path{doi:10.1103/PhysRevLett.74.4763}}.

\bibitem{Kofman1996}
A.~G. Kofman, G.~Kurizki, Quantum zeno effect on atomic excitation decay in
  resonators, Phys. Rev. A 54 (1996) R3750--R3753.
\newblock \href {http://dx.doi.org/10.1103/PhysRevA.54.R3750}
  {\path{doi:10.1103/PhysRevA.54.R3750}}.

\bibitem{Zheng2013}
W.~Zheng, D.~Z. Xu, X.~Peng, X.~Zhou, J.~Du, C.~P. Sun, Experimental
  demonstration of the quantum zeno effect in nmr with entanglement-based
  measurements, Phys. Rev. A 87 (2013) 032112.
\newblock \href {http://dx.doi.org/10.1103/PhysRevA.87.032112}
  {\path{doi:10.1103/PhysRevA.87.032112}}.

\bibitem{Wolters2013}
J.~Wolters, M.~Strau\ss{}, R.~S. Schoenfeld, O.~Benson, Quantum zeno phenomenon
  on a single solid-state spin, Phys. Rev. A 88 (2013) 020101,
  arXiv:1301.4544v2 [quant-ph].
\newblock \href {http://dx.doi.org/10.1103/PhysRevA.88.020101}
  {\path{doi:10.1103/PhysRevA.88.020101}}.

\bibitem{Fischer2001}
M.~C. Fischer, B.~Guti\'errez-Medina, M.~G. Raizen, Observation of the quantum
  zeno and anti-zeno effects in an unstable system, Phys. Rev. Lett. 87 (2001)
  040402.
\newblock \href {http://dx.doi.org/10.1103/PhysRevLett.87.040402}
  {\path{doi:10.1103/PhysRevLett.87.040402}}.

\bibitem{Facchi2001}
P.~Facchi, H.~Nakazato, S.~Pascazio, From the quantum zeno to the inverse
  quantum zeno effect, Phys. Rev. Lett. 86 (2001) 2699--2703,
  arXiv:quant-ph/0006094v2.
\newblock \href {http://dx.doi.org/10.1103/PhysRevLett.86.2699}
  {\path{doi:10.1103/PhysRevLett.86.2699}}.

\bibitem{Steinberg1993}
A.~M. Steinberg, P.~G. Kwiat, R.~Y. Chiao, Measurement of the single-photon
  tunneling time, Phys. Rev. Lett. 71 (1993) 708--711.
\newblock \href {http://dx.doi.org/10.1103/PhysRevLett.71.708}
  {\path{doi:10.1103/PhysRevLett.71.708}}.

\end{thebibliography}

\end{document}